\documentclass[12pt]{article}
\usepackage{amssymb,amsmath,amsthm,amsfonts,mathrsfs}
\usepackage{enumerate}
\usepackage{ragged2e}
\usepackage{pgf,tikz}
\usetikzlibrary{intersections}

\usepackage{ucs} 
\usepackage[utf8x]{inputenc}

\usepackage{tikz}
\usepackage{tkz-tab}
\usepackage{caption}
\usepackage{latexsym}
\usepackage{amssymb}
\usepackage{amsmath}
\usepackage{subcaption} 
\usepackage{graphicx}
\usepackage{pgfplots}
\pgfplotsset{compat=1.15}
\usepackage{mathrsfs}
\usetikzlibrary{arrows}
\usepackage{float}
\usepackage{amsmath}
\usepackage[colorlinks=true, allcolors=blue]{hyperref}
\usepackage[all]{hypcap}
\usepackage[capitalize,nameinlink]{cleveref}
\usepackage{titlesec}
\titlelabel{\thetitle.\quad}
\usepackage{float}
\usepackage{pgfplots}
\pgfplotsset{compat=1.15}
\usepackage{mathrsfs}
\usetikzlibrary{arrows}

\newtheorem{theo}{Theorem}

\newtheorem{lem}[theo]{Lemma}

\theoremstyle{definition}
\newtheorem{case}{Case}[theo]

\usepackage{amsmath}

\newcommand*{\QEDA}{\hfill\ensuremath{\blacksquare}}
\usepackage{ragged2e}

\begin{document}
\title{Signed interval graphs and bigraphs: A generalization of interval graphs and bigraphs}
\def\correspondingauthor{\footnote{Corresponding author}}
\author{Ashok Kumar Das\correspondingauthor{}  and  Indrajit Paul\\Department of Pure Mathematics\\
University of Calcutta\\
35, Ballygunge Circular Road\\
Kolkata-700019\\
Email Address - ashokdas.cu@gmail.com \&
paulindrajit199822@gmail.com}
\maketitle
\begin{abstract}
    In this paper, we define and characterize signed interval graphs and bigraphs  introducing the concept of negative interval. Also we have shown that these classes of graphs are respectively a generalization of well known classes of interval graphs and interval bigraphs. In this context we have observed that signed interval graphs coincide with the complement of Threshold tolerance graphs(co-TT graphs) introduced by Monma, Reed and Trotter \cite{22}. Finally, we have solved the open problem of forbidden induced subgraph characterization of co-TT graphs posed by them in the same paper. 
\end{abstract}
\section{Introduction}
A graph $G= (V, E)$ is an \textit{interval graph} if corresponding to each vertex $v \in V$ we can assign an interval of the real line such that two vertices are adjacent if and only if their corresponding intervals intersect.  Interval graphs are important in graph theory. They have several elegant characterizations and efficient recognition algorithms \cite{1,3,4,11,12,13,18,19,21,24}
 Also has found application in various real world problem.\par

An interval digraph is a \textit{directed graph} representable by assigning each vertex $v$ an ordered pair $(S_v, T_v)$ of closed intervals so that $uv$ is a (directed) edge if and only if $S_u$ intersects $T_v$. The sets $S_v$ and $T_v$ are the \textit{source set} and the \textit{sink set} for $v$. A bipartite graph (in short, bigraph) $B = (X, Y, E)$ is an interval bigraph if there exists a one to one correspondence between the vertex set $X \cup Y$ of $B$ and a collection of intervals $\{I_v: v \in X \cup Y\}$ so that $xy \in E$, if and only if $I_x \cap I_y \neq \emptyset$.\par

The \textit{biadjacency matrix} of a bipartite graph is the submatrix of its adjacency matrix consisting of the rows indexed by the vertices of one partite set and columns by the vertices of another. Interval digraphs and interval bigraphs were introduced in \cite{25} and \cite{16} respectively and as observed in \cite{6}, the two concepts are equivalent.\par

The point is that the adjacency matrix of an interval digraph is the bi-adjacency matrix of an
interval bigraph and conversely the bi-adjacency matrix of an interval bigraph becomes the adjacency matrix
of an interval digraph,by adding, if necessary rows or columns of $0$s to make it square.\par

Several characterization of interval bigraphs are known (see~\cite{17,25,27}). One characterization uses \textit{Ferrers bigraph} ( introduced independently by Guttman \cite{15} and Riguet \cite{23} ), which are satisfying any of the following equivalent conditions,
\begin{enumerate}[i)]
\item The set of neighbors of any partite sets are linearly ordered by inclusion.
\item The rows and the columns of the biadjacency matrix can be permuted independently, so that the $1$'s cluster in the upper right (or, lower left) as a Ferrers diagram.
\item The biadjacency matrix  has no $2$-by-$2$ permutation matrix $\left( \begin{array}{ c  c } $1$ & $0$ \\ $0$ & $1$ \end{array} \right) or \left( \begin{array}{ c  c } $0$ & $1$ \\ $1$ & $0$ \end{array} \right)$ as a submatrix.
\end{enumerate}

A \emph{zero-partition} of a binary matrix is a coloring of each $0$ with $R$ or $C$ in such a way that every $R$ has only $0$s colored $R$ to its right and every $C$ has only $0$s colored $C$ below it. A matrix that admits a zero partition after suitable row and column permutation is zero-partitionable. Following theorem characterizes interval bigraphs.\par

\begin{theo}[Sen et al.~\cite{25}]
The following conditions are equivalent.
\begin{enumerate}[(a)]
\item B is an interval bigraph.
\item The biadjacency matrix of B is zero partitionable.
\item B is the intersection of two Ferrers bigraphs whose union is a complete.
\end{enumerate}
\end{theo}

Any binary matrix (that is a $(0,1)$ - matrix) with only one zero is a Ferrers matrix. Hence every bigraph is the intersection of finite number of  Ferrers Bigraphs whose intersection is $B$. The minimum number of Ferrers bigraphs whose intersection is $B$ is the \textit{Ferrers dimension} of the bigraph $B$, written $fdim(B)$.\par

By Theorem~1, every interval bigraph has Ferrers dimension at most $2$. But the converse is not true \cite{25}. Bigraphs with Ferrers dimension $2$ were characterized independently by Cogis \cite{2} and others. Cogis \cite{2} introduced the \textit{associated graph} $H(B)$ for a bigraph $B$. Its vertices are the $0$'s of its biadjacency matrix of $B$, with two such vertices are adjacent in $H(B)$ if and only if they are the $0$'s of the $2$-by-$2$ permutation matrix. Cogis \cite{2} proved that $fdim(B) \leq 2$ if and only if $H(B)$ is bipartite. Sen et al. translate Cogis's condition to an adjacency matrix condition for a bigraph $B$ to be of $fdim(B) \leq 2$ in the following theorem.\par

\begin{theo}[\cite{25}]
A bigraph B has Ferrers dimension at most $2$ if and only if the rows and the columns of the biadjacency matrix $A(B)$ of $B$ can be permuted independently, so that no 0 has a 1 both below it and to its right.
\end{theo}
\begin{math}
$\subsection{Motivation}$
\end{math}
\indent\hspace{.05in}Let $G= (V, E)$ be an interval graph and $\{I_v : v \in V\}$ be its interval representation. Assume $l(v)$ and $r(v)$ respectively denote the left end point and right end point of $I_v$. It can be observed that two intervals $I_u$ and $I_v$ intersect if and only if left end point of each interval is less than or equal to the right end point of the other. Thus $I_u \cap I_v \neq \emptyset$ if and only if $l(u) \leq r(v)$ and $l(v) \leq r(u)$.\par
Here $l(v)$ and $r(v)$ respectively denote the left end point and right end point of $I_v$. Again for $I_v$ to be an interval it is needed that $l(v) \leq r(v)$. Now, if we relax this condition for $I_v$ and we assume it is possible that $l(v) > r(v)$, then we call such an interval a negative interval and denote it by ${I_v}^-$. In the other case where $l(v) \leq r(v)$ in $I_v$, we say $I_v$ is a positive interval and denote it by ${I_v}^+$.\par

The motivation of the concept of negative interval comes from the definition of complement of Threshold tolerance graph \cite{22}. A graph $G$ is a \textit{Threshold tolerance graph} if corresponding to each vertex $v$ we can assign a positive weight $w_v$ and a positive tolerance $t_v$ so that $uv$ is an edge if and only if $w_u + w_v$ $>$ $min\{t_u, t_v\}$. The complements of threshold tolerance graphs are called \textit{co-TT graphs}. Threshold tolerance graphs are threshold graphs when we assume all the tolerances to be equal. In other words threshold tolerance graphs are the generalization of threshold graphs. The following characterization of co-TT graphs is due to Monma, Reed, and Trotter \cite{22}.\par

\begin{theo}
A graph G is a co-TT graph if and only if we can assign positive numbers $a_v$ and $b_v$ for each $v \in V$ such that,
 $$xy \in E(G) \Leftrightarrow a_x \leq b_y \text{ and } a_y \leq b_x \hspace{1cm}\ldots (1)$$
\end{theo}

For a co-TT graph $G$ the above condition is equivalent to assignment of a pair of positive numbers $(a_v, b_v)$ to each vertex $v$ of $G$, where it is possible that $a_v \leq b_v$ or $a_v > b_v$. If $a_v \leq b_v$, we say that the pair of positive numbers $(a_v, b_v)$ forms a positive interval ${I_v}^+ = [a_v, b_v]$ and in the other case if $a_v > b_v$ then we say that the pair $(a_v, b_v)$ forms a negative interval ${I_v}^- = [a_v, b_v]$. It is easy to observe that when two intervals ${I_x}^+ = [a_x, b_x]$ and ${I_y}^+ = [a_y, b_y]$ intersect then the condition $(1)$ is satisfied. Also if one of the two intervals, say ${I_x}^-$ (or, ${I_y}^-$) is a negative interval and ${I_x}^- \subseteq {I_y}^+$ $(or {I_y}^- \subseteq {I_x}^+ )$ then also the condition $(1)$ is satisfied. Thus we are in a position to give a general definition of interval graphs which we call signed interval graphs. A graph $G = (V, E)$ is a \textit{signed interval graph} if corresponding to each vertex $v$ we assign either positive interval ${I_v}^+$ or negative interval ${I_v}^-$ such that $uv \in V(E)$ if and only if ${I_u}^+ \cap {I_v}^+ \neq \emptyset$ or ${I_v}^- \subseteq {I_u}^+$ or ${I_u}^- \subseteq {I_v}^+$.\par

 When all the intervals $I_v$ are positive in the representation we have the interval graph. Also from the above observation we conclude that the signed interval graphs coincide with co-TT graphs.

\section{Characterization}
Now, we extend the definition of signed interval graphs to the class of bipartite graphs. A bigraph $B = (X, Y, E)$ is a signed interval bigraph, if corresponding to each vertex $v \in X \cup Y$ we can assign a positive interval or negative interval such that $xy \in E$ if and only if ${I_x}^+ \cap {I_y}^+ \neq \emptyset$ or ${I_x}^- \subseteq {I_y}^+$ or ${I_y}^- \subseteq {I_x}^+$ .\par
In the following theorem we characterize signed interval bigraphs.
\begin{theo}
For a bigraph $B= (X, Y, E)$ the following conditions are equivalent.
\begin{enumerate}[(a)]
\item $B$ is a signed interval bigraph.
\item $B$ is a bigraph of Ferrers dimension at most $2$.
\end{enumerate}
\end{theo}

\noindent \textbf{Proof.}  $(a) \Rightarrow (b)$ Let $B$ be a signed interval bigraph. Without loss of generality we may assume that all the left end point of the intervals are distinct in the representation of $B$. Now, we arrange the rows and columns of $A(B)$ respectively according to the increasing order of the left end points of $I_{x_k}$, $x_k \in X$ and $I_{y_l}$, $y_l \in Y$. Next, suppose that $(i, j)$th entry of $A(B)$ is zero. We consider the following cases.
\begin{case}
Suppose ${I_{x_i}}^+ \cap {I_{y_j}}^+ = \emptyset$. Then we have $r(x_i) < l(y_j)$ or $r(y_j) < l(x_i)$. The first possibility implies for all ${I_{{y_k}}}^+$ $(k>j)$, ${I_{x_i}}^+ \cap {I_{y_k}}^+ = \emptyset$ and for all ${I_{{y_k}}}^-$ $(k>j)$, $ {I_{y_k}}^- \nsubseteq {I_{x_i}}^+ $. Thus $x_iy_k = 0$ for all $k \geqslant j$. In other words all the entries to the right of $(i, j)$th entry are $0$s. Again, if $r(y_j) < l(x_i)$, then for all ${I_{{x_k}}}^+$ $(k>i)$, ${I_{x_i}}^+ \cap {I_{y_j}}^+ = \emptyset$. Also for all ${I_{{x_k}}}^-$ $(k>i)$, $ {I_{x_k}}^- \nsubseteq {I_{y_j}}^+ $. So all the entries are zeros below $(i, j)$th entry.
\end{case}
\begin{case}
Suppose ${{I_y}_j}^- \nsubseteq {{I_x}_i}^+$. Now we have the following possibilities: either $r(y_j) < l(x_i)$ or $l(y_j) > r(x_i)$. In the first possibility, for all $x_k$ $(k > i)$, $ {I_{y_j}}^- \nsubseteq {I_{x_k}}^+ $. So all the entries are zero below $(i, j)$th entry. In the other possibility, for all $y_k$ $(k>j)$, ${I_{y_k}}^+ \cap {I_{x_i}}^+ = \emptyset$ or $ {I_{y_k}}^- \nsubseteq {I_{x_i}}^+ $. Thus all the entries to the right of $(i, j)$th entry are $0$s.
\end{case}
\begin{case}
Suppose ${{I_x}_i}^- \nsubseteq {{I_y}_j}^+$. Then we have the following possibilities: either $l(x_i) < r(y_j)$ or $l(y_j) < r(x_i)$. As before, in the first possibility all the entries to the right of $(i, j)$th entry are $0$s and in the second possibility all the entries below $(i, j)$th entry are $0$s.
\end{case}
Thus, with this arrangement of rows and columns of $A(B)$, the bigraph $B$ is of Ferrers dimension $2$.\par

$(b) \Rightarrow (a)$ Let $B = (X, Y, E)$ be a bigraph of Ferrers dimension $2$. Also suppose that the vertices can be arranged as $x_1$, $x_2$, \ldots , $x_n$ and $y_1$, $y_2$, \ldots , $y_m$ respectively in the rows and columns of the biadjacency matrix $A(B)$ of $B$ such that no $0$ has a $1$ both below and to its right, where $x_i \in X$ $(1 \leq i \leq n)$ and $y_j \in Y$ $(1 \leq j \leq m)$.\par

Next, we assign intervals to the vertices of $X$ as follows. Let $x_i$ be any vertex of $X$ and in the $x_i$th row suppose the last $1$ appears in the $k$ th column. Then the interval corresponding to $x_i$ is $[i, k]$ and we denote it by ${{I_x}_i}^+$ if $i \leq k$ or ${{I_x}_i}^-$ if $i > k$. In the similar way we assign an interval $[j, l]$ corresponding to the vertex $y_j$ . It will be denoted by ${{I_y}_j}^+$ if $j \leq l$ or ${{I_y}_j}^-$ if $j > l$.\par

Now, we shall prove that with this interval assignment, $B$ is a signed interval bigraph. Let $x_iy_j \in E$, i.e. $(i, j)$th entry of $A(B)$ is $1$. Let the interval corresponding to $x_i$ is $[i, k]$ and  the interval corresponding to $y_j$ is $[j, l]$. Again $(i, j)$th entry of is $1$ implies $j \leq k$ and $i \leq l$. Now consider the three following cases.\par \vspace{.2cm}
\noindent\textbf{Case~1.\ }Suppose ${{I_x}_i}$ and ${{I_y}_j}$ are both positive intervals. Then  $j \leq k$ and $i \leq l$ imply ${I_{x_i}}^+ \cap {I_{y_j}}^+ \neq \emptyset$.\par \vspace{.2cm}
\noindent\textbf{Case~2.\ }Suppose ${{I_x}_i}$ be a positive interval and ${{I_y}_j}$ be a negative interval. Then  $i \leq k$ and $j > l$. Also $j \leq k$ and $i \leq l$ imply ${I_{y_j}}^-  \subseteq {I_{x_i}}^+$.\par \vspace{.2cm}
\noindent\textbf{Case~3.\ }Suppose ${{I_x}_i}$ be a negative interval and ${{I_y}_j}$ be a positive interval. Then  $j \leq l$ and $i>k$. Again $j \leq k$ and $i \leq l$ imply ${I_{x_i}}^-  \subseteq {I_{y_j}}^+$.\par \vspace{.2cm}

Next, suppose that $(i, j)$th entry is $0$ and all the entries to its right are $0$s. Then $k < j$ and let $l > i$. Consider the following possibilities. If ${{I_x}_i}$ and ${{I_y}_j}$ are both positive intervals, then  $k < j$ implies ${I_{x_i}}^+ \cap {I_{y_j}}^+ = \emptyset$. Again, if ${{I_x}_i}$ is a negative interval and ${{I_y}_j}$ is a positive interval, then $k < j$ implies ${I_{x_i}}^-  \nsubseteq {I_{y_j}}^+$. Next if ${{I_x}_i}$ is positive interval and ${{I_y}_j}$ is a negative interval, then again $k < j$ implies ${I_{y_j}}^-  \nsubseteq {I_{x_i}}^+$.\par

Finally, suppose that $(i, j)$th entry is $0$ and all the entries are $0$s below it. Then we can assume that $k > j$ and $l < i$. Then as before we can show that ${I_{x_i}}^+ \cap {I_{y_j}}^+ = \emptyset$. Also if one of the two intervals, say ${{I_x}_i}$ is a negative interval then ${I_{x_i}}^-  \nsubseteq {I_{y_j}}^+$. This completes the proof of the theorem.\QEDA

Thus signed interval bigraphs are a generalization of interval bigraphs. Interval bigraphs are those signed interval bigraphs where all the vertices are assigned positive intervals.\par

In the next theorem we give an analogous characterization of signed interval graphs, introduced earlier, in terms of their adjacency matrices. We assume that signed interval graphs have self loops corresponding to the vertices where we assign positive intervals. Thus among the diagonal entries of the adjacency matrix of a signed interval graph we have a $1$ corresponding to the vertices where positive intervals are assigned and the others are $0$s. \par

\begin{theo}
For a graph G= (V, E) the following conditions are equivalent.
\begin{enumerate}[(a)]
\item G is a signed interval graph.
\item The adjacency matrix $A(G)$ of $G$ is of Ferrers dimension 2, where all the diagonal entries of $A(G)$ are not all $1$ or $0$s. 
\end{enumerate}
\end{theo}
\noindent\textbf{Proof.} $(a) \Rightarrow (b)$ Let G be a signed interval graph. Without loss of generality we may assume that the left end point of the intervals assigned to the vertex set $V$ are distinct. We arrange the rows(columns) of the adjacency matrix $A = A(G)$ of $G$ according to the increasing order of the left end points of the corresponding intervals. Next, we shall show that the matrix $A$ is such that if the $(i, j)$th entry of $A$ is a zero then each entry to the right of it is also a zero or each entry below it is also zero.\par
Now consider the following cases.
\begin{case}
Suppose ${I_{v_i}}^+ \cap {I_{v_j}}^+ = \emptyset$ and $i < j$, then $r(v_i) < l(v_j)$. Now $l(v_k) > r(v_i)$, $\forall\ k \geq j$. This implies that all the entries to the right of $(i, j)$th entry are also $0$s. Also all the entries to the below of $(j, i)$th entry are $0$s. Next, assume ${I_{v_i}}^+ \cap {I_{v_j}}^+ =\emptyset$ and $i>j$, then $r(v_j)<l(v_i)$. Similarly we conclude that all the entries below $(i,j)$th entry are $0$s and all the entries to the right of $(j,i)$th entry are $0$s.
\end{case}
\begin{case}
Let ${I_{v_j}}^- \nsubseteq {I_{v_i}}^+ $ and $i < j$, then we have either $l(v_j) > r(v_i)$ or $r(v_j) < l(v_i)$. Then in the first possibility we have $l(v_k) > r(v_i)$, $\forall k \geq j$. So all the entries to the right of $(i, j)$th entry are $0$s and all the entries below $(j,i)$th entry are $0$s. In the other possibility for all $k \geq i$, ${I_{v_j}}^- \nsubseteq {I_{v_k}}^+ $. Thus all the entries below $(i, j)$th entry are $0$s and all the entries to the right of $(j,i)$the entry are $0$s.
\end{case}
\begin{case}
Let ${I_{v_i}}^- \nsubseteq {I_{v_j}}^+ $ and $j < i$, then we have either $r(v_i) < l(v_j)$ or $r(v_j) < l(v_i)$. In the first possibility $\forall k > j$, ${I_{v_i}}^- \nsubseteq {I_{v_k}}^+ $ and hence all the entries to the right of $(i, j)$th entry are $0$s. Also $(j, i)$th entry is a zero and all the entries below $(j, i)$th entry are $0$s. Similarly in the second possibility we can show that all the entries to the right of $(j, i)$th entry are $0$s and all the entries below $(i, j)$th entry are $0$s.
\end{case}
It can be easily observed that the matrix $A$ is symmetric. Without loss of generality may assume the $(i, j)$th entry is $1$. Then we have either ${I_{v_i}}^+ \cap {I_{v_j}}^+ \neq \emptyset$ or ${I_{v_i}}^- \subseteq {I_{v_j}}^+$ or ${I_{v_j}}^- \subseteq {I_{v_i}}^+ $. In any case $(j, i)$th entry is also a  $1$ and if the $(i, j)$th entry of $A$ is $0$, then we have  ${I_{v_i}}^+ \cap {I_{v_j}}^+ = \emptyset$ or ${I_{v_j}}^- \nsubseteq {I_{v_i}}^+ $ or  ${I_{v_i}}^- \nsubseteq {I_{v_j}}^+ $. Then also $(j, i)$th entry of $A$ is also $0$.\par
Now by Theorem~$2$, $A = A(G)$ is a matrix of Ferrers dimension $2$, where all the diagonal entries are not $1$ or $0$s as the intervals are not all positive or negative. \par
$(b) \Rightarrow (a)$ Let the adjacency matrix $A=A(G)$ of $G$ is of Ferrers dimension $2$, where all diagonal entries are not $1$ or $0$s.Thus we can arrange the rows(columns) of $A$ such that no $0$ has a $1$ both to its right and below it. Now, we can construct a signed interval representation of $G$ as described in the Theorem $4$.\QEDA \vspace{.5cm}

\section{Forbidden induced subgraphs}
A graph $G$ is a \textit{split graph} if its vertex set can be partitioned into a stable set and a complete graph.\\
Golumbic et al.\cite{14} have given the forbidden induced subgraphs of split co-TT graphs (which are both split graph and co-TT graphs)using the work of Trotter and Moore\cite{26}.\\
\textit{Circular arc graph} is the intersection graph of a family of circular arcs of a host circle. If the vertices of the circular arc graph can be covered by two disjoint cliques then it is a \textit{two-clique circular arc graph}. Trotter and Moore \cite{26} characterized two clique circular arc graphs in terms of forbidden induced subgraphs. They present the forbidden families as a set system and proved that $G$ is a two-clique circular arc graph if and only if it's complement $ \overline{G}$ contains no induced subgraphs of the form $G_1,G_2,G_3$ and several infinite families $C_i,T_i,W_i,D_i,M_i,N_i $ $(i\geq1)$ (see Fig.1).\\
\begin{figure}[H]
$\mathcal{C}_3 = \{\{1,2\},\{2,3\},\{3,1\}\}$ \\
$\mathcal{C}_4 = \{\{1,2\},\{2,3\},\{3,4\},\{4,1\}\}$ \\
$\mathcal{C}_5 = \{\{1,2\},\{2,3\},\{3,4\},\{4,5\},\{5,1\}\}$ \\
\dots \\
$\mathcal{T}_1 = \{\{1,2\},\{2,3\},\{3,4\},\{2,3,5\},\{5\}\}$ \\
$\mathcal{T}_2 = \{\{1,2\},\{2,3\},\{3,4\},\{4,5\},\{2,3,4,6\},\{6\}\}$ \\
$\mathcal{T}_3 = \{\{1,2\},\{2,3\},\{3,4\},\{4,5\},\{5,6\},\{2,3,4,5,7\},\{7\}\}$ \\
\dots \\
$\mathcal{W}_1 = \{\{1,2\},\{2,3\},\{1,2,4\},\{2,3,4\},\{4\}\}$ \\
$\mathcal{W}_2 = \{\{1,2\},\{2,3\},\{3,4\},\{1,2,3,5\},\{2,3,4,5\},\{5\}\}$ \\
$\mathcal{W}_3 = \{\{1,2\},\{2,3\},\{3,4\},\{4,5\},\{1,2,3,4,6\},\{2,3,4,5,6\},\{6\}\}$ \\
\dots \\
$\mathcal{D}_1 = \{\{1,2,5\},\{2,3,5\},\{3\},\{4,5\},\{2,3,4,5\}\}$ \\
$\mathcal{D}_2 = \{\{1,2,6\},\{2,3,6\},\{3,4,6\},\{4\},\{5,6\},\{2,3,4,5,6\}\}$ \\
$\mathcal{D}_3 =\{\{1,2,7\},\{2,3,7\},\{3,4,7\},\{4,5,7\},\{5\},\{6,7\},\{2,3,4,5,6,7\}\}$ \\
\dots \\
$\mathcal{M}_1 = \{\{1,2,3,4,5\},\{1,2,3\},\{1\},\{1,2,4,6\},\{2,4\},\{2,5\}\}$ \\
$\mathcal{M}_2 = \{\{1,2,3,4,5,6,7\},\{1,2,3,4,5\},\{1,2,3\},\{1\},\{1,2,3,4,6,8\},\{1,2,4,6\},\{2,4\},\{2,7\}\}$ \\
$\mathcal{M}_3 =\{\{1,2,3,4,5,6,7,8,9\},\{1,2,3,4,5,6,7\},\{1,2,3,4,5\},\{1,2,3\},\{1\}, \{1,2,3,4,5,6,8,10\},$

\hfill
$\{1,2,3,4,6,8\},\{1,2,4,6\},\{2,4\},\{2,9\}\}$ \\
\dots \\
$\mathcal{N}_1 = \{\{1,2,3\},\{1\},\{1,2,4,6\},\{2,4\},\{2,5\},\{6\}\}$ \\
$\mathcal{N}_2 = \{\{1,2,3,4,5\},\{1,2,3\},\{1\},\{1,2,3,4,6,8\},\{1,2,4,6\},\{2,4\},\{2,7\},\{8\}\}$ \\
$\mathcal{N}_3 = \{\{1,2,3,4,5,6,7\},\{1,2,3,4,5\},\{1,2,3\},\{1\},\{1,2,3,4,5,6,8,10\},\{1,2,3,4,6,8\},$

\hfill 
$\{1,2,4,6\},\{2,4\},\{2,9\},\{10\}\}$ \\
\dots \\
$\mathcal{G}_1 = \{\{1,3,5\},\{1,2\},\{3,4\},\{5,6\}\}$ \\
$\mathcal{G}_2 = \{\{1\},\{1,2,3,4\},\{2,4,5\},\{2,3,6\}\}$ \\
$\mathcal{G}_3 = \{\{1,2\},\{3,4\},\{5\},\{1,2,3\},\{1,3,5\}\}$\\
 \caption*{Fig.1: Forbidden families of two-clique circular arc graphs as in \cite{26}.}
    \label{}
\end{figure}

\par Feder,Hell and Huang \cite{10} explained how to obtain the forbidden bipartite graphs from the Fig.1. These bigraphs are presented in Fig.2.\\
\begin{figure}[H]
\centering
\begin{minipage}{.5\textwidth}
  \centering
    \begin{tikzpicture}[scale=.7]
\tikzstyle{every node}=[circle, draw, fill=black,
                        inner sep=1pt, minimum width=2.5pt]
 \node (a1) at (1,3){};
    \node(a2) at(1,4){};
    \node(a3) at (2,3){};
    \node(a4) at (2,4){};
    \node(a5) at (1,2){};
    \node(a6) at (2,2){};
    \node(a7) at(1,1){};
    \node(a8) at (2,1){};
    \node(a9) at (0,2){};
    \node(a10) at (0,3){};
    \node(a11) at(3,3){};
    \node(a12) at (3,2){};
    \draw (a1)--(a2);
    \draw (a3)--(a4);\draw(a2)--(a4);
    \draw(a1)--(a3);
    \draw(a3)--(a11);
    \draw(a11)--(a12);
    \draw(a12)--(a6);
    \draw(a6)--(a5);
    \draw(a5)--(a9);
    \draw(a9)--(a10);
    \draw(a10)--(a1);
    \draw(a1)--(a12);
    \draw(a3)--(a9);
    \draw(a1)--(a5);
    \draw(a3)--(a6);
    \draw(a6)--(a8);
    \draw(a5)--(a7);
    
    \end{tikzpicture}\\
   $\mathcal{M}_1 $
\end{minipage}%
\begin{minipage}{.5\textwidth}
  \centering
 \begin{tikzpicture}[scale=.7]
\tikzstyle{every node}=[circle, draw, fill=black,
                        inner sep=1pt, minimum width=2.5pt]
  \node(a) at(7,7){};
   \node(b) at (8,7){};
   \node(c) at (7,8){};
   \node(d) at (8,8){};
   \node(e) at (5,6){};
   \node(f) at (5,5){};
   \node(g) at (5,4){};
   \node(h) at (5,3){};
   \node(i) at (7,2){};
   \node(j) at (8,2){};
   \node(k) at (10,3){};
   \node(l) at (10,4){};
   \node(m) at (10,5){};
   \node(n) at (10,6){};
   \node(o) at (7,1){};
   \node(p) at (8,1){};
   \draw(a)--(b);
   \draw(a)--(c);
   \draw(c)--(d);
   \draw(d)--(b);
   \draw(a)--(e);
   \draw(e)--(f);
   \draw(f)--(g);
   \draw(g)--(h);
   \draw(h)--(i);
   \draw(i)--(j);
   \draw(j)--(k);
   \draw(k)--(l);
   \draw(l)--(m);
   \draw(m)--(n);
   \draw(n)--(b);
   \draw(i)--(o);
   \draw(j)--(p);
   \draw(a)--(g);
   \draw(a)--(i);
   \draw(a)--(k);
   \draw(a)--(m);
   \draw(b)--(f);
   \draw(b)--(h);
   \draw(b)--(j);
   \draw(b)--(l);
   \draw(f)--(i);
   \draw(g)--(j);
   \draw(i)--(l);
   \draw(j)--(m);
  \draw[dashed](11,4.5)--(13,4.5);
  
    \end{tikzpicture}\\
   $\mathcal{M}_2 $
\end{minipage}

\end{figure}
\begin{figure}[H]
\centering
\begin{minipage}{.5\textwidth}
  \centering
    \begin{tikzpicture}[scale=.7]
\tikzstyle{every node}=[circle, draw, fill=black,
                        inner sep=1pt, minimum width=2.5pt]
\node(a) at (0,1){};
\node(b) at (1,1){};
\node(c) at (2,1){};
\node(d) at (2,0){};
\node(e) at (3,0){};
\node(f) at (3,1){};
\node(g) at (4,1){};
\node(h) at (5,1){};
\node(i) at (2,2){};
\node(j) at (2,3){};
\node(k) at (3,2){};
\node(l) at (3,3){};
\draw(a)--(b);
\draw(b)--(c);
\draw(c)--(d);
\draw(d)--(e);
\draw(e)--(f);
\draw(f)--(g);
\draw(g)--(h);
\draw(c)--(f);
\draw(c)--(i);
\draw(i)--(j);
\draw(i)--(k);
\draw(k)--(l);
\draw(f)--(k);
 
    \end{tikzpicture}\\
   $\mathcal{N}_1 $
\end{minipage}%
\begin{minipage}{.5\textwidth}
  \centering
 \begin{tikzpicture}[scale=.7]
\tikzstyle{every node}=[circle, draw, fill=black,
                        inner sep=1pt, minimum width=2.5pt]
\node(a) at (5,6){};
\node(b) at (6,6){}; 
\node(c) at (4,5){};
\node(d) at (4,4){};
\node(e) at (4,3){};
\node(f) at (5,2){};
\node(g) at (6,2){};
\node(h) at (7,3){};
\node(i) at (7,4){};
\node(j) at (7,5){};
\node(k) at (8,6){};
\node(l) at (3,6){};
\node(m) at (8,4){};
\node(n) at (9,4){};
\node(o) at (3,4){};
\node(p) at (2,4){};
\draw(a)--(b);
\draw(a)--(c);
\draw(c)--(d);
\draw(c)--(l);
\draw(d)--(o);
\draw(o)--(p);
\draw(d)--(e);
\draw(e)--(f);
\draw(f)--(g);
\draw(g)--(h);
\draw(h)--(i);
\draw(i)--(m);
\draw(m)--(n);
\draw(i)--(j);
\draw(j)--(k);
\draw(j)--(b);
\draw(a)--(i);
\draw(b)--(d);
\draw(c)--(j);
\draw(c)--(f);
\draw(d)--(i);
\draw(d)--(g);
\draw(f)--(i);
\draw(g)--(j);
\draw[dashed](10,4)--(12,4);
    \end{tikzpicture}\\
   $\mathcal{N}_2 $
\end{minipage}

\end{figure}
\begin{figure}[H]
    \centering
    \begin{tikzpicture}[scale=1]
\tikzstyle{every node}=[circle, draw, fill=black,
                        inner sep=0pt, minimum width=2.5pt]
\node(a) at (4,5){};
\node(b) at (5,5){};
\node(c) at (6,4){};
\node(d) at (6,3){};
\node(e) at (5,2){};
\node(f) at (4,2){};
\node(g) at (3,3){};
\node(h) at (3,4){};
\draw (g)--(h);
\draw(h)--(a);
\draw(a)--(b);
\draw(b)--(c);
\draw(c)--(d);
\draw(d)--(e);
\draw(e)--(f);
\draw[dashed](f)--(g);
   \end{tikzpicture}\\
    $\mathcal{C}_i $
\end{figure}

\begin{figure}[H]
\centering
\begin{tikzpicture}[scale=1]
{\tikzstyle{every node}=[circle, draw, fill=black,
                        inner sep=0pt, minimum width=2.5pt]
\draw (0:0) node(a){}--++(100:1)node(b){}--++(150:1)node(c){}--++(100:-1)node(d){}--cycle;
\draw (a)--++(0:.6)node{}--++(0:.6)node(e){};
\draw[dashed] (e)--++(0:1)node(f){};
\draw (f)--++(80:1)node(g){}--++(30:1)node(h){}--++(80:-1)node(i){}--++(30:-1) (b)--(e) (b)--(f) (g)--(a) (g)--(e);
\path [name path=l] (b) --++ (30:2);
\path [name path=m] (g) --++ (-30:-2);
\path [name intersections={of=l and m, by=j}];
\draw (b)--(j)node{}--(g) (j)--++(90:.7)node(k){};
\draw[line width=.4mm] (c)--(d) (h)--(i) (j)--(k);

\coordinate (a1) at ([xshift=2.4cm]f);
\draw (a1)node{}--++(0:.6)node(b1){}--++(90:1)node(c1){}--++(0:1.6)node(d1){}--++(90:1)node(e1){}--++(0:-1.6)node(f1){}--++(90:-1) (b1)--++(0:.6)node(g1){}--(d1)--(a1);
\draw[dashed] (g1)--++(0:1.6)node(h1){};
\draw (h1)--(c1);
\path[name path=i1] (a1)--++(100:1);
\path[name path=j1] (c1)--++(30:-1);
\path[name intersections={of=i1 and j1, by=k1}];
\path[name path=l1] (h1)--++(80:1);
\path[name path=m1] (d1)--++(-30:1);
\path[name intersections={of=l1 and m1, by=n1}];
\draw (a1)--(k1)node{}--(c1) (k1)--++(110:1)node(o1){};
\draw (h1)--(n1)node{}--(d1) (n1)--++(70:1)node(p1){};
\draw[line width=.4mm] (k1)--(o1) (n1)--(p1) (e1)--(f1);
\coordinate (z1) at ([xshift=.8cm]b1);

\coordinate (a2) at ([xshift=1.2cm]h1);
\draw (a2)node{}--++(0:.6)node(b2){}--++(0:.6)node(c2){}--++(0:.6)node{}--++(0:.6)node(d2){};
\draw[dashed] (d2)--++(0:1)node(e2){};
\draw (e2)--++(0:.6)node(f2){}--++(0:.6)node(g2){};
\path[name path=h2] (c2)--++(45:2);
\path[name path=i2] (e2)--++(-45:-2);
\path[name intersections={of=h2 and i2, by=j2}];
\draw (c2)--(j2)node{}--(e2) (j2)--(d2) (j2)--++(90:.6)node(k2){}--++(90:.6)node(l2){};
\draw[line width=.4mm] (a2)--(b2) (f2)--(g2) (k2)--(l2);

\coordinate (a3) at ([yshift=-3.6cm]d);
\draw (a3)node{}--++(0:.6)node(b3){}--++(0:.6)node{}--++(0:.6)node(c3){}--++(90:.7)node{}--++(90:.7)node(d3){}--++(90:.7)node(e3){} (c3)--++(0:.6)node{}--++(0:.6)node(f3){}--++(0:.6)node(g3){};
\draw[line width=.4mm] (a3)--(b3) (e3)--(d3) (f3)--(g3);

\coordinate (a4) at ([xshift=1.5cm]g3);
\draw (a4)node{}--++(0:.6)node(b4){}--++(0:.6)node(c4){}--++(0:1.2)node(d4){}--++(0:.6)node(e4){}--++(0:.6)node(f4){} (c4)--++(90:1)node{}--++(90:1)node(g4){}--++(0:1.2)node(h4){}--++(90:-1)node{}--++(0:-1.2) (d4)--++(90:1);
\draw[line width=.4mm] (a4)--(b4) (e4)--(f4) (g4)--(h4);
\coordinate (z2) at ([xshift=.6cm]c4);

\coordinate (a5) at ([xshift=1.5cm]f4);
\draw (a5)node{}--++(0:.7)node(b5){}--++(90:.8)node{}--++(0:.9)node(c5){}--++(0:.9)node{}--++(90:-.8)node(d5){}--++(0:.7)node(e5){} (b5)--++(0:.9)node(f5){}--(e5) (f5)--(c5)--++(90:.7)node(g5){}--++(90:.7)node(h5){};
\draw[line width=.4mm] (a5)--(b5) (d5)--(e5) (g5)--(h5);
}

\coordinate (z) at (0,-0.5);
\draw (z-|j)node{$\mathcal{W}_i$} (z-|z1)node{$\mathcal{D}_i$} (z-|j2)node{$\mathcal{T}_i$};
\coordinate (y) at ([yshift=-.5cm]c3);
\draw (y)node{$\mathcal{G}_1$} (y-|z2)node{$\mathcal{G}_3$} (y-|c5)node{$\mathcal{G}_2$};
\end{tikzpicture}\\ \vspace{.5cm}
Fig 2: Forbidden bigraphs from Fig 1.\\
\end{figure}\vspace{-.4cm}

Das,Sen and others obtained the above  forbidden graphs (except the family $N_i$) in different ways. In a bipartite graph $B$, three edges $e_1,e_2$, and $e_3$ form an \textit{asteroidal triple of edges}(ATE) if there is a path joining two edges that avoids the neighbors of the third. Das and Sen\cite{7} showed that a bigraph B having Ferrars dimension 2 is ATE-free. A bigraph is called \textit{chordal bipartite} or \textit{bichordal} if it does not contain any cycle of length greater than 4. Das and Sen\cite{8} also determined the minimal set of bigraphs that are bichordal and contains an ATE. Which are the graphs $G_1,G_2,G_3$ and the infinite families $T_i,W_i,D_i$. Das and Chakraborty\cite{5} have determined the family $M_i$ as forbidden family for bigraphs of Ferrers dimension 2. It can be observed that the family $C_i$ is the family of cycles $C_{2n}$ $(n\geq3)$.\\
Now we will use the Tortter and Moore's graphs of Fig.2 to obtain the forbidden induced subgraphs of signed interval bigraphs and signed interval graphs. First, we state a result of J.Huang \cite{20} connecting two-clique circular arc graphs and the bigraphs of Ferrers dimension 2.\\

\begin{theo}[\cite{20}]
A graph is a two clique circular arc graph if and only if its complement $\overline{G}$ is a bigraph of Ferrers dimension at most 2.
\end{theo}\par
Therefore the bigraphs given in Fig.2 are the forbidden induced subgraphs of the bigraphs of Ferrers dimension 2 and hence from Theorem 4 they are also the forbidden induced subgraphs of signed interval bigraphs.\\
Golumbic,Weingarten and Limouzy \cite{14} obtained the forbidden induced subgraphs of split co-TT graphs from the graphs of Fig.2 in the following way.\\
Let $B=(X,Y,E)$ be any bigraph of Fig.2. Then the graph $G$ whose vertex set $V(G)$ is $X\cup Y$ and the edge set $E(G)$ is $E\cup X\times X$ or $E\cup Y\times Y$
is a split graph and forbidden graph for split co-TT graphs. Then removing the isomorphic graphs they have obtained the complete list of forbidden induced subgraphs for split co-TT graphs. Obviously these graphs are also forbidden induced subgraphs for co-TT graphs. The following forbidden induced subgraph (Fig.3) of split co-TT graph is obtained from the graph $C_3$ of Fig.1 as describe above.
\begin{figure}[H]
\centering
\begin{tikzpicture}[scale=1.5]
\tikzstyle{every node}=[circle, draw, fill=black,
                        inner sep=2pt, minimum width=2.5pt]
\node(a1) at (0,0){};
\node(a2) at (2,2){};
\node(a3) at (4,0){};
\node(a4) at (1,1){};
\node(a5) at (3,1){};
\node(a6) at (2,0){};
\draw (a1)--(a2);
\draw(a2)--(a3);
\draw(a1)--(a3);
\draw (a4)--(a5);
\draw (a5)--(a6);
\draw (a4)--(a6);
\end{tikzpicture}\\ \vspace{.5cm}
   Fig 3: Forbidden induced subgraph of split co-TT graph
\end{figure}
\noindent Already we have observed that the signed interval graphs coincide with co-TT graphs. Now using the result of Theorem 5, we can state the following theorem. 
\begin{theo}
The following are equivalent 
\begin{enumerate}[(i)]
    \item $G$ is a co-TT graph.
    \item $G$ is a signed interval graph.
    \item The adjacency matrix $A(G)$ of $G$ is of Ferrers dimension 2, where all diagonal entries are not all $1s$ or $0s$.
\end{enumerate}

\end{theo}\par 
 The above Theorem 7 also provides a representation characterization of co-TT graphs, the problem posed by Monma, Reed and Trotter\cite{22}.\\
Now before presenting the forbidden families of co-TT graph, we first state some definitions from the literature. A graph $G$ is \textit{chordal} if it does not contain any chordless cycle $C_n$ $(n\geq 4)$. The strongly chordal graphs are introduced by Faber \cite{9} and satisfies several equivalent conditions. Here we state the forbidden subgraph characterization of strongly chordal graphs. A \textit{trampoline} is a graph $G$ formed from an even cycle $v_1,v_2,v_3,...,v_{2k},v_1$ by adding an edge between even subscripted vertices so that the vertices ${v_2,v_4,...v_{2k}}$ induce a complete subgraph. The graph of Fig.3 is a trampoline with $k=3$. The following graph in Fig.4 is a trampoline with $k=4$. A trampoline with $2k$ vertices is also called $k$-\textit{sun}.\\
\begin{figure}[H]
\centering
\begin{tikzpicture}[scale=1]
\tikzstyle{every node}=[circle, draw, fill=black,
                        inner sep=2pt, minimum width=2.5pt]
\node(a1) at (4,3){};
\node(a2) at (2,3){};
\node(a3) at (2,1){};
\node(a4) at (4,1){};
\node(a5) at (5,2){};
\node(a6) at (3,0){};
\node(a7) at (1,2){};
\node(a8) at (3,4){};
\draw (a1)--(a2);
\draw(a2)--(a3);
\draw(a3)--(a4);
\draw (a4)--(a1);
\draw (a5)--(a4);
\draw (a4)--(a6);
\draw (a1)--(a5);
\draw (a3)--(a6);
\draw (a3)--(a7);
\draw (a7)--(a2);
\draw (a2)--(a8);
\draw (a8)--(a1);
\draw (a1)--(a3);
\draw (a2)--(a4);
\end{tikzpicture}\\ \vspace{.5cm}
   Fig 4: The Sun graph $S_4$ 
\end{figure}
\begin{theo}[\cite{9}]
A chordal graph $G$ is strongly chordal graph if and only if $G$ does not contain a trampoline as an induced subgraph .
\end{theo}
Now it follows from \cite{22} that every co-TT graph is strongly chordal. This implies that every co-TT graph is chordal and the sun graphs $S_k $ $(k\geq 3)$ are also forbidden induced subgraphs for co-TT graphs.\\
\begin{figure}[H]
    \begin{subfigure}[b]{0.32\textwidth}
    \centering
    \resizebox{\linewidth}{!}{
    \begin{tikzpicture}[scale=1]
\tikzstyle{every node}=[circle, draw, fill=black,
                        inner sep=1.5pt, minimum width=2.5pt]
\draw(0,0) node[label={[label distance=1pt]180:${\scriptstyle }$}] {};
\draw(1,1) node[label={[label distance=1pt]135:${\scriptstyle }$}] {};
\draw(2,2) node[label={[label distance=1pt]0:${\scriptstyle }$}] {};
\draw(2,3) node[label={[label distance=1pt]0:${\scriptstyle }$}] {};
\draw(2,4) node[label={[label distance=1pt]0:${\scriptstyle }$}] {};
\draw(3,1) node[label={[label distance=1pt]0:${\scriptstyle }$}] {};
\draw(4,0) node[label={[label distance=1pt]0:${\scriptstyle }$}] {};
\draw(0,0)--(1,1);
\draw(1,1)--(2,2);
\draw(2,2)--(2,3);
\draw(2,3)--(2,4);
\draw(2,2)--(3,1);
\draw(3,1)--(4,0);
\end{tikzpicture} 
}
$(a)$ $T$
    \end{subfigure}
\begin{subfigure}[b]{0.32\textwidth}
\centering
\resizebox{\linewidth}{!}{
 \begin{tikzpicture}[scale=1]
\tikzstyle{every node}=[circle, draw, fill=black,
                        inner sep=1.5pt, minimum width=2.5pt]
 \draw(0,0) node[label={[label distance=1pt]-90:${\scriptstyle }$}] {};
 \draw(1,0) node[label={[label distance=1pt]-90:${\scriptstyle }$}] {};
 \draw(2,0) node[label={[label distance=1pt]135:${\scriptstyle }$}] {};
 \draw(4,0) node[label={[label distance=1pt]45:${\scriptstyle }$}] {};
 \draw(5,0) node[label={[label distance=1pt]-90:${\scriptstyle }$}] {};
 \draw(6,0) node[label={[label distance=1pt]-90:${\scriptstyle }$}] {};
 \draw(3,2) node[label={[label distance=1pt]0:${\scriptstyle }$}] {};
 \draw(3,3) node[label={[label distance=1pt]0:${\scriptstyle }$}] {};
 \draw(3,4) node[label={[label distance=1pt]0:${\scriptstyle }$}] {};
 \draw(0,0)--(1,0);
 \draw(1,0)--(2,0);
 \draw(2,0)--(4,0);
 \draw(4,0)--(5,0);
 \draw(5,0)--(6,0);
 \draw(2,0)--(3,2);
 \draw(3,2)--(4,0);
 \draw(3,2)--(3,3);
 \draw(3,3)--(3,4);
 \end{tikzpicture}
 }
 $(b)$ $T_0$
 
\end{subfigure}
\begin{subfigure}[b]{0.32\textwidth}
\centering
\resizebox{\linewidth}{!}{
 \begin{tikzpicture}[scale=1]
\tikzstyle{every node}=[circle, draw, fill=black,
                        inner sep=1pt, minimum width=2.5pt]
 
 \draw(1,0) node[label={[label distance=1pt]-90:${\scriptstyle }$}] {};
 \draw(2,0) node[label={[label distance=1pt]135:${\scriptstyle }$}] {};
 \draw(4,0) node[label={[label distance=1pt]45:${\scriptstyle }$}] {};
 \draw(5,0) node[label={[label distance=1pt]-90:${\scriptstyle }$}] {};
  \draw(3,0) node[label={[label distance=1pt]-90:${\scriptstyle }$}] {};
 \draw(3,1) node[label={[label distance=1pt]0:${\scriptstyle }$}] {};
 \draw(3,2) node[label={[label distance=1pt]0:${\scriptstyle }$}] {};
 \draw(3,3) node[label={[label distance=1pt]0:${\scriptstyle }$}] {};
 \draw(3,2)--(3,0);
 
 \draw(1,0)--(2,0);
 \draw(2,0)--(4,0);
 \draw(4,0)--(5,0);
 
 \draw(2,0)--(3,1);
 \draw(3,1)--(4,0);
 \draw(3,1)--(3,2);
 \draw(3,2)--(3,3);
 \end{tikzpicture}
 }
 $(c)$  $P$
 
\end{subfigure}\vspace{.2cm}
\centering
Fig 5: The forbidden graphs $T$, $T_0$ and $P$.

\end{figure}

\noindent In \cite{22} Monma,Reed and Tortter have given two forbidden induced subgraph Fig 5$(a)$ and $(b)$ (without proof) for co-TT graph. In the next two lemma we shall use  Theorem 7 to show that they are forbidden induced subgraphs for co-TT graphs.\vspace{.2cm}
\begin{figure}[H]
    \begin{subfigure}[b]{0.32\textwidth}
    \centering
    \resizebox{\linewidth}{!}{
    \begin{tikzpicture}[scale=1]
\tikzstyle{every node}=[circle, draw, fill=black,
                        inner sep=1.5pt, minimum width=2.5pt]
\draw(0,0) node[label={[label distance=1pt,scale=1.8]180:${\scriptstyle z}$}] {};
\draw(1,1) node[label={[label distance=1pt,scale=1.8]135:${\scriptstyle w}$}] {};
\draw(2,2) node[label={[label distance=1pt,scale=1.8]0:${\scriptstyle s}$}] {};
\draw(2,3) node[label={[label distance=1pt,scale=1.8]0:${\scriptstyle u}$}] {};
\draw(2,4) node[label={[label distance=1pt,scale=1.8]0:${\scriptstyle x}$}] {};
\draw(3,1) node[label={[label distance=1pt,scale=1.8]0:${\scriptstyle v}$}] {};
\draw(4,0) node[label={[label distance=1pt,scale=1.8]0:${\scriptstyle y}$}] {};
\draw(0,0)--(1,1);
\draw(1,1)--(2,2);
\draw(2,2)--(2,3);
\draw(2,3)--(2,4);
\draw(2,2)--(3,1);
\draw(3,1)--(4,0);
\end{tikzpicture} 
}
    \end{subfigure}
\begin{subfigure}[b]{0.32\textwidth}
\centering
\resizebox{\linewidth}{!}{
 \begin{tikzpicture}[scale=1]
\tikzstyle{every node}=[circle, draw, fill=black,
                        inner sep=1.5pt, minimum width=2.5pt]
 \draw(0,0) node[label={[label distance=1pt,scale=2.2]-90:${\scriptstyle q}$}] {};
 \draw(1,0) node[label={[label distance=1pt,scale=2.2]-90:${\scriptstyle y}$}] {};
 \draw(2,0) node[label={[label distance=1pt,scale=2.2]135:${\scriptstyle v}$}] {};
 \draw(4,0) node[label={[label distance=1pt,scale=2.2]45:${\scriptstyle w}$}] {};
 \draw(5,0) node[label={[label distance=1pt,scale=2.2]-90:${\scriptstyle z}$}] {};
 \draw(6,0) node[label={[label distance=1pt,scale=2.2]-90:${\scriptstyle p}$}] {};
 \draw(3,2) node[label={[label distance=1pt,scale=2.2]0:${\scriptstyle u}$}] {};
 \draw(3,3) node[label={[label distance=1pt,scale=2.2]0:${\scriptstyle x}$}] {};
 \draw(3,4) node[label={[label distance=1pt,scale=2.2]0:${\scriptstyle s}$}] {};
 \draw(0,0)--(1,0);
 \draw(1,0)--(2,0);
 \draw(2,0)--(4,0);
 \draw(4,0)--(5,0);
 \draw(5,0)--(6,0);
 \draw(2,0)--(3,2);
 \draw(3,2)--(4,0);
 \draw(3,2)--(3,3);
 \draw(3,3)--(3,4);
 \end{tikzpicture}
 }
 
\end{subfigure}
\begin{subfigure}[b]{0.32\textwidth}
\centering
\resizebox{\linewidth}{!}{
 \begin{tikzpicture}[scale=1]
\tikzstyle{every node}=[circle, draw, fill=black,
                        inner sep=1pt, minimum width=2.5pt]
 
 \draw(1,0) node[label={[label distance=1pt,scale=1.5]-90:${\scriptstyle y}$}] {};
 \draw(2,0) node[label={[label distance=1pt,scale=1.5]135:${\scriptstyle v}$}] {};
 \draw(4,0) node[label={[label distance=1pt,scale=1.5]45:${\scriptstyle w}$}] {};
 \draw(5,0) node[label={[label distance=1pt,scale=1.5]-90:${\scriptstyle z}$}] {};
  \draw(3,0) node[label={[label distance=1pt,scale=1.5]-90:${\scriptstyle t}$}] {};
 \draw(3,1) node[label={[label distance=1pt,scale=1.5]0:${\scriptstyle u}$}] {};
 \draw(3,2) node[label={[label distance=1pt,scale=1.5]0:${\scriptstyle x}$}] {};
 \draw(3,3) node[label={[label distance=1pt,scale=1.5]0:${\scriptstyle s}$}] {};
 \draw(3,2)--(3,0);
 
 \draw(1,0)--(2,0);
 \draw(2,0)--(4,0);
 \draw(4,0)--(5,0);
 
 \draw(2,0)--(3,1);
 \draw(3,1)--(4,0);
 \draw(3,1)--(3,2);
 \draw(3,2)--(3,3);
 \end{tikzpicture}
 }

\end{subfigure}\vspace{.5cm}
\centering
Fig 6: A labeling of the graphs $T$, $T_0$ and $P$.

\end{figure}

\begin{lem}
The graph $T$ of Fig.$5(a)$ is a forbidden induced graph for co-TT graphs.
\end{lem}
\noindent\textbf{Proof.} We shall show that the graph $T$ has no signed interval representation. In the graph $T$, the vertices $u,v$ and $w$ are adjacent to the vertex $s$ but no two of them are adjacent. Take all the intervals $I_s,I_u,I_v$ and $I_w$ as positive intervals, where any two of the intervals $I_u,I_v$ and $I_w$ are disjoint and all the intervals intersects $I_s$. Also let $l(u) < l(s)$ and $r(s) < r(w)$. And $I_v$ is contained in $I_s$, where $r(u) < l(v)$ and $r(v) < l(w)$. Now we can take positive or negative intervals for the vertices $x$ and $z$ such that $xu$ and $wz$ are edges but $xs$ and $zs$ are non edges. However, if we take any positive or negative interval for $y$ such that $yv$ is an edge then $ys$ is also an edge. Thus $T$ has no signed interval representation and accordingly $T$ is a forbidden induced subgraph for co-TT graphs.\QEDA
\begin{lem}
The graph $T_0$ of Fig.$5(b)$ is a forbidden induced graph for co-TT graph.
\end{lem}
\noindent\textbf{Proof.} As in the Lemma 9, we shall show that the graph $T_0$ has no signed interval representation. Since the graph $T_0$ is symmetric and the vertices $u,v$ and $w$ form a clique, all the positive intervals $I_u,I_v$ and $I_w$ intersect. Also, assume $l(u) < l(v) < l(w)$ and $r(u) < r(v) < r(w)$. Next, assume $I_x$ and $I_z$ are positive intervals such that $I_x$ intersects $I_u$ only and $I_z$ intersects $I_w$ only. Also, we can take a negative interval $I^{-}_y$ such that $I^{-}_y$ is contained in $I_v$ only. Now assume $l(x) < l(u)$ and $r(w) < r(z)$. Then we can take positive or negative intervals for $s$ and $p$ such that $xs$ and $zp$ are edges (i.e. $s$ is adjacent to $x$ only and $p$ is adjacent to $z$ only). But we can't have any positive or negative interval for $q$ such that $q$ is adjacent to $y$ only. Thus $T_0$ has no signed interval representation and this completes the proof.\QEDA 
\begin{lem}
The graph $P$ is a forbidden induced subgraph for co-TT graphs.
\end{lem}
\noindent\textbf{Proof.} As before we shall show that $P$ has no signed interval representation. Again the graph $P$ is symmetric and the vertices $u,v,t$ and $u,w,t$ form two cliques. Without loss of generality we consider the positive intervals $I_u,I_v$ and $I_t$ for the vertices $u$,$v$ and $t$ respectively such that these intervals intersect and $l(v)<l(u)<l(t)$ and $r(v)<r(u)<r(t)$. Next consider a positive interval $I_w$ such that $I_w$ intersects $I_u$ and $I_t$ where $r(v)<l(w)$ and $r(t)<r(w)$. Now we take positive intervals $I_y$ and $I_z$ such that $I_y$ intersects $I_v$ only and $I_z$ intersects $I_w$ only. Next, we take a negative interval $I^-_x$ such that $I^-_x$ contained in $I_u$ only, i.e. $l(u)<r(x)<l(x)$ and $r(v)<l(x)<r(t)$. Then there is no choice for the interval $I_s$ (positive or negative) such that $sx$ is the only edge of $P$. This completes the proof of the lemma.\QEDA
\begin{figure}[H]
    \centering
 \begin{tikzpicture}[scale=1]
\tikzstyle{every node}=[circle, draw, fill=black,
                        inner sep=2pt, minimum width=2.5pt]
 \node(a) at (0,0){};
 \node(b) at (1,0){};
 \node(c) at (2,0){};
 \node(d) at (3,0){};
 \node(e) at (5,0){};
 \node(f) at (6,0){};
 \node(g) at (3,2){};
 \node(h) at (3,3){};
 \node(i) at (3,4){};
 \draw(a)--(b);
 \draw(b)--(c);
 \draw(c)--(d);
 \draw(e)--(f);
 \draw(b)--(g);
 \draw(c)--(g);
 \draw(d)--(g);
 \draw(e)--(g);
 \draw(h)--(g);
 \draw(h)--(i);
 \draw[dashed](d)--(e);

   \end{tikzpicture}\\ \vspace{.5cm}
   Fig 7: The family of graphs $P_i(i\geq 2)$
\end{figure}
\noindent In the next lemma we shall prove that the family of graphs $P_i(i\geq 2)$ is a forbidden family of graphs for the co-TT graphs.
\begin{figure}[H]
    \centering
 \begin{tikzpicture}[scale=1]
\tikzstyle{every node}=[circle, draw, fill=black,
                        inner sep=2pt, minimum width=2.5pt]
\draw(0,0) node[label={[label distance=1pt,scale=1.5]-90:${\scriptstyle y}$}] {};
\draw(1,0) node[label={[label distance=1pt,scale=1.5]90:${\scriptstyle v}$}] {};
\draw(2,0) node[label={[label distance=1pt,scale=1.5]-90:${\scriptstyle t_1}$}] {};
\draw(3,0) node[label={[label distance=1pt,scale=1.5]-90:${\scriptstyle t_2}$}] {};
\draw(4,0) node[label={[label distance=1pt,scale=1.5]-90:${\scriptstyle t_i}$}] {};

\draw(5,0) node[label={[label distance=1pt,scale=1.5]90:${\scriptstyle w}$}] {};
\draw(6,0) node[label={[label distance=1pt,scale=1.5]90:${\scriptstyle z}$}] {};
\draw(3,2) node[label={[label distance=1pt,scale=1.5]0:${\scriptstyle u}$}] {};
 \draw(3,3) node[label={[label distance=1pt,scale=1.5]0:${\scriptstyle x}$}] {};
  \draw(3,4) node[label={[label distance=1pt,scale=1.5]0:${\scriptstyle s}$}] {};
\draw(0,0)--(1,0);
\draw(1,0)--(2,0);
\draw(2,0)--(3,0);
\draw[dashed](3,0)--(4,0);
\draw(5,0)--(6,0);
\draw(1,0)--(3,2);
\draw(3,2)--(5,0);
\draw(3,3)--(3,2);
\draw(3,3)--(3,4);
\draw(2,0)--(3,2);
\draw(3,0)--(3,2);
\draw(4,0)--(5,0);
\draw(3,2)--(4,0);
   \end{tikzpicture}\\ \vspace{.5cm}
   Fig 8: A labeling of the graph $P_i(i\geq 2)$
\end{figure}
\begin{lem}
The family of graphs $P_i(i\geq2)$ is a forbidden family of graphs for the co-TT graphs.
\end{lem}
\noindent\textbf{Proof.} We shall show that the graph $P_i$ has no signed interval representation. As in the Lemma 11 we take the positive intervals $I_v,I_u,I_{t_1},I_{t_2},...,I_{t_i},I_w$ such that 
\begin{center}
    $l(v)<l(u)<l(t_1)<l(t_2)<...<l(t_{i-1})<l(t_i)<l(w)$,\\ 
    and $r(v)<r(t_1)<r(t_2)<...<r(t_{i-1})<r(u)<r(t_i)<r(w)$.
\end{center}
Also $r(v)<l(t_2)$, $r(t_{k-2})<l(t_k)$ for ($k\geq3$) and $r(t_{i-1})<l(w)$. Next, we take the intervals $I^+_y$, $I^+_z$ and $I^-_x$ such that $I^+_y$ intersects $I_v$ only and $I^+_z$ intersects $I_w$ only. Also $l(u)<r(x)$ and $r(v)<l(x)<r(t_1)$, then $I^-_x$ contained in $I_u$ only but we can't have an interval $I_s$ for the vertex $s$, such that $s$ is adjacent to $x$ only in $P_i$. This completes the proof.\QEDA
\par
Now we address the problem of forbidden induced subgraph characterization of co-TT graphs, the open problem posed by Monma, Reed and Trotter\cite{22}, using the bigraphs of Fig.2.

First, we note that every co-TT graph must be chordal and the graph $\mathcal{G}_1$ contains $T$ as an induced subgraph. Next, we add minimum number of edges among the vertices of any partite set of each families $\mathcal{C}_i$, $\mathcal{T}_i$, $\mathcal{D}_i$, $\mathcal{W}_i$, $\mathcal{M}_i$, $\mathcal{N}_i$ of bigraphs to make them chordal graphs. Let the families of graphs so obtain are respectively $\mathcal{C'}_i$, $\mathcal{T'}_i$, $\mathcal{W'}_i$, $\mathcal{D'}_i$, $\mathcal{M'}_i$, and $\mathcal{N'}_i$, after removing the isomorphic graphs and the graphs which contain the previously mentioned forbidden graphs as an induced subgraph. Also the bigraphs $\mathcal{G}_2$ and $\mathcal{G}_3$ contain $T$ as an induced subgraph after making them chordal. Next, we observe that every graph of the family $\mathcal{C'}_i$ contains the graph $S_3$ as an induced subgraph. Also every graph of the family $\mathcal{T'}_i$ contains a graph of the family $P_i (i\geq 2)$ as an induced subgraph. Again let $S$ be the forbidden family for split co-TT graphs (as describe in \cite{14}). Which are also forbidden induced subgraphs of co-TT graphs.  
\par considering all the observations, we obtain the following characterization of co-TT graphs in terms of forbidden induced subgraphs.
\begin{theo}
A graph $G$ is a co-TT graph if and only if $G$ does not contain any graph of the families $S_k(k\geq 3)$, $S-\{S_3\}$, $\mathcal{W'}_i$, $\mathcal{D'}_i$, $\mathcal{M'}_i$, $\mathcal{N'}_i$, $P_i(i\geq2)$ and the graphs $T$, $T_0$ and $P$ as an induced subgraph. 
\end{theo}

\end{document}